\documentclass[conference]{IEEEtran}
\IEEEoverridecommandlockouts
\usepackage{amsmath,amssymb,amsfonts}
\usepackage{algorithmic}
\usepackage{graphicx}
\usepackage{textcomp}
\usepackage{xcolor}
\def\BibTeX{{\rm B\kern-.05em{\sc i\kern-.025em b}\kern-.08em
    T\kern-.1667em\lower.7ex\hbox{E}\kern-.125emX}}
\begin{document}

\title{Face Verification Using 60~GHz 802.11 waveforms}

\author{Eran Hof \and Amichai Sanderovich \and Evyatar Hemo \\ Qualcomm Israel Ltd\\P.O. Box 1212 \\ Israel \\ ehof, amichais, ehemo  @qti.qualcomm.com}

\author{\IEEEauthorblockN{Eran Hof}
\IEEEauthorblockA{\textit{Qualcomm Israel Ltd} \\
P.O. Box 1212 \\ Israel \\ ehof@qti.qualcomm.com}
\and
\IEEEauthorblockN{Amichai Sanderovich}
\IEEEauthorblockA{\textit{Qualcomm Israel Ltd} \\
P.O. Box 1212 \\ Israel \\ amichais@qti.qualcomm.com}   
\and
\IEEEauthorblockN{Evyatar Hemo}
\IEEEauthorblockA{\textit{Qualcomm Israel Ltd} \\
P.O. Box 1212 \\ Israel \\ ehemo@qti.qualcomm.com}}

\maketitle

\begin{abstract}
Verification of an identity based on the human face radar signature in mmwave is studied. The chipset for 802.11 Qualcomm® ad/y networking that is cable of operating in a radar mode is used. A dataset with faces of 200 different persons was collected for the testing. Our preliminary study shows promising results for the application of autoencoder for the setup at hand. 
\end{abstract}


\section{Introduction}

Automatic biometric authentication is becoming increasingly popular as mean of identification in front of machines \cite{wayman2005introduction}. The main advantage of such authentication is, of course, removing the need for remembering passwords or pass phrases. This allows a non-expert and occasional user to securely operate machines with full authentication. %First examples for such usages were in defense and security systems where finger-print or iris scan were used in order to identify the person in front of the machine. In recent years, and especially as smart-phones became the digital extension of the physical person, these techniques were deployed in mass scale and are now widely used. 
Face verification (see \cite{1467368},\cite{abate20072d} and also \cite{tran2017regressing}) is a recent important addition to the biometric arsenal techniques. It is non-intrusive, hands-free, and is acceptable by most users.   
%Verification is the only biometric authentication we are testing in this paper. However, it is noted that additional usages for biometric authentications may be provided by the hardware setting studied in this work and are the possible focus of future research (in particular, classification where the machine needs to decide who is the person and duplicate detect where the system recognizes that the person was not registered before, are two promising applications of the chipset hardware studied in this work).

Different sensors are used for face recognition. RGB camera, is one of the well known sensors. The main drawback of the RGB camera is that it suffers from variable lighting condition. It addition, RGB camera suffers from poor-detection performance of a mask or photograph (see \cite{wayman2005introduction}). In order to overcome these, RGB-D (D for depth) sensors are used for facial verification. These sensors include structured light sensors and time-of-flight sensors (see, e.g.,  \cite{3DAssistedFaceRecognition2005},\cite{Bakirman17},\cite{luna2017robust}). In this work, we study the potential of face verification based only on the radar signature as captured by the Millimeter-wave (mmWave) networking schipset system that can be operated in a radar fashion. The chipset used in our study if a fully functional 802.11 Qualcomm®\footnote{Qualcomm 802.11 is a product of Qualcomm Technologies, Inc. and/or it’s subsidiaries.} ad/y networking communication (see, e.g.,~\cite{ABIResearch60Ghz, smulders2002exploiting}). Recent interests in mmWave radars is found in gesture recognition and related applications, see e.g.~\cite{molchanov2015short,lien2016soli,yeo2016radarcat} and references therein. 
Scanners for security applications based on mmWave technology are adopted in airports worldwide and ongoing research effort focus on advanced algorithmic technique for analyzing mmWave signals, see, e.g., \cite{harmer2012review,sakamoto2016fast,gonzalez2016millimeter,lopez2018using, yujiri2003passive, ahmed2012advanced, noujeim2014compact, feger2013low, oppelt2017mimo}.

A 802.11ad/y packet starts with a short training field, followed by a channel estimation field (CEF), packet header, physical layer (PHY) payload and optional fields for gain control and additional training. %The PHY-payload in turn is composed from MAC header, MAC payload and error detection coding. 
A CEF is composed of Golay complementary sequences (128 symbols long) used in estimating the channel response characteristics. Complementary Golay sequences are well-studied signals in the radar community (see, e.g., \cite{haderer2014comparison,levanon2017complementary, pezeshki2008doppler}). Emerging 802.11ad technology and its PHY suitability for radar applications motivated the study in providing opportunistic radar devices based on 802.11ad technology~\cite{kumari2015investigating,kumari2017ieee,grossi2018opportunistic}. This is somehow different then legacy frequency modulated continuous wave (FMCW) radars (see, e.g., \cite{wenger1998automotive,hasch2012millimeter}). %The key differences and relative advantages of using Golay complementary sequences signals instead of FMCW is summarized in Section~\ref{sec:FMCWVsGolay}.
%
%In this paper we study facial verification using a 802.11 ad/y networking chipset that can be operated as a radar. In particular we operate the chipset with two RF chains, one for transmission and one as a receiver. These two RF chips are operated simultaneously to provide radar capabilities. 
%By operating the chipset in this way we are able to capture the unique radar signature of each face. The face verification is done taking advantage of the massive-antenna elements available in the chipset (the massive number of antenna elements (64) is used to overcome high pathloss for WiFi links) so that the radar signature is detailed enough.
%
Our observations for the facial verification are the Golay correlation outputs (see \cite{nitsche2014ieee}) for each of the transmit/receive antenna element pairs.  Golay sequences enable us to remove non-relevant reflections and focus the verification processing on returning wave only from the face surface. To the best of our knowledge, no such dataset is publicly available. %We therefore captured the faces with a real 802.11ad chipset \cite{dataset2}.
%As the number of antenna elements is smaller, the verification quality is worse, since the radar signature is less detailed.

The rest of this paper continues as follows, in Section~\ref{sec:SystemDescription} the radar system and related technology are described, the capturing process and the dataset building is provided in Section \ref{sec:capture},
the approach we used for the verification is detailed in Section~\ref{sec:FaceVerification} and Section~\ref{sec:conclus} concludes the paper. 

\section{Sensor Description} \label{sec:SystemDescription}
%In this section we describe the capturing system: its capabilities and the main performance metrics. 
%\subsection{General Description}

\begin{figure}[!t]
\centering
\includegraphics[width=6.0cm]{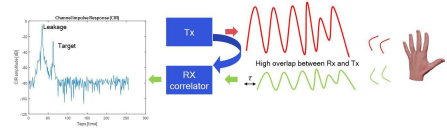}
\caption{A Radar system with a target and the corresponding received signal.} \label{fig:LeakageAndTarget}
\end{figure}

\begin{figure}[!t]
\centering
\includegraphics[width=6.0cm]{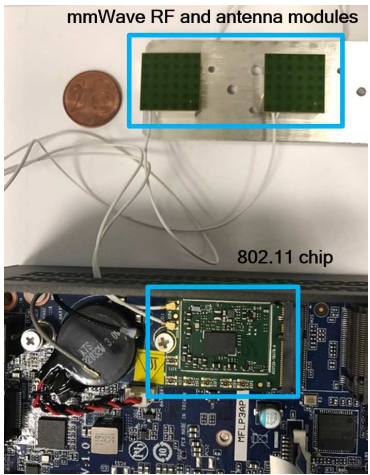}
\caption{A Radar setup based on Qualcomm's 802.11~ad/ay communication system. The system includes a 802.11 chip connected to two RF and antenna modules. The system is highly miniature in size as compared to a 2 euro-cent coin.} \label{fig:RMChips}
\end{figure}

An illustration of a radar system is depicted in Fig.~\ref{fig:LeakageAndTarget}. An electromagnetic wave is transmitted from the radar tx module and reflected back from a target object (a hand in Fig.~\ref{fig:LeakageAndTarget}). Some electromagnetic energy is reflected back to the location of the receiver which can sample the received signal and detect the presence of a target. %By estimating the time-of-flight and the angle-of-arrival, the radar device can estimate the location and the speed of the the target at hand. 
Our sensor, depicted in Fig.~\ref{fig:RMChips}, is similar to the radar scheme described in~\cite{kumari2015investigating}, we are re-using an existing communication system as a 60~GHz radar sensor. 60~GHz is favorable due to the small wavelegth (5mm), the high bandwidth (13GHz) and the small antenna size (2.5mm). %Moreover, since the system was built for 802.11 high speed communication, it includes low-cost high availability CMOS chipset, with mass market support. 

%\begin{figure}

%The digital signal consists of re-using the Golay sequences from 802.11ad/ay modem ~\cite{ghasempour2017ieee} for the radar operation of transmission and reception processing. 
%The sensor is connected to a computer using PCIe link, allowing processing using MATLAB and Python.

\section{Face Capturing Using mmWave Radar}\label{sec:capture}

\begin{figure}
	\centering
		\includegraphics[width=6.0cm]{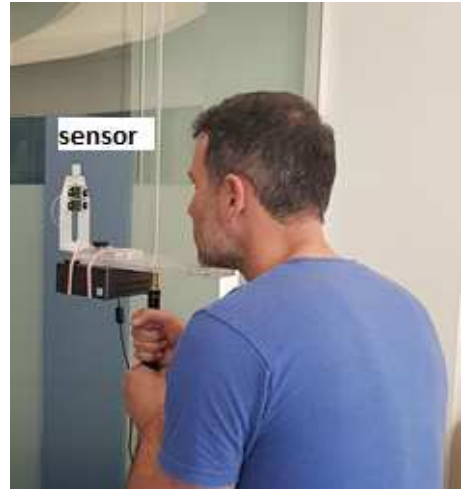}
	\caption{The capturing setup in action}
	\label{fig:test}
\end{figure}

We first describe the capturing of the dataset used in our study. The head of each subject is placed in front of the radar sensor, for two bursts, each taking 3 seconds with about 8 seconds in between. During this time, the sensor captured 200 frames per orientation/distance. %We used a plastic fixture for all subjects so that the location of the faces will remain the same for all the dataset. 
The signal is recorded in time corresponding to distances of 30cm and 50cm. In addition, capturing is repeated for for different orientations: 24,15 degrees left and right, and looking at the center. The capturing is demonstrated in Fig. \ref{fig:test}.%, where the sensor faces the subject, along the plastic fixture.
%We used a plastic fixture and not metal to reduce its effect on the radar signature. 
The subjects in our study included both men and women, at variable ages, some were wearing glasses and some having beards. For each pair of transmit and receive antenna elements (32 TX $\times$ 32 RX antenna element pairs for the chipset at hand), the Golay correlation outputs for the distances at hand (and the corresponding depsth, e.g. 24 cm for a human face) are recorded. Unlike camera image, each of these complex-valued numbers represents the energy that was reflected from the face. That is, there is no ''empty'' pixels. %This is why we believe that such capture contains enough distinctive information between the faces.
%The scanned dataset can be found online at \cite{dataset2}.

\section{Face Verification} \label{sec:FaceVerification}

\begin{figure}[!t]
\centering
\includegraphics[width=6.0cm]{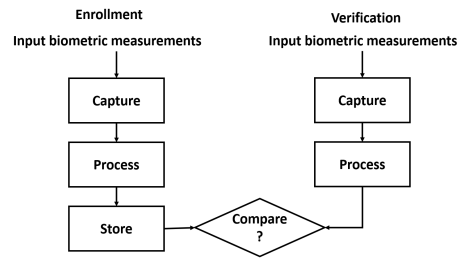}
\caption{A two stage biometric verification process} \label{fig:biometricVerification}
\end{figure}

The studied authentication scheme is based on a two stage process as detailed in Fig. \ref{fig:biometricVerification}: enrollment of the person where the training data is captured, and then a verification query that the registered person is indeed standing in front of the radar by comparing the captured frame to the training data during the enrollment stage. 

%\subsection{One-Class vs Multi-Class Classification} \label{sec:oneclass}

We distinguish between one-class and multi-class classificaiotn. A multi-class classification is a problem where an unknown image needs to be classified into a class out of several possible classes. For example, by classifying apples and oranges, we can determine if the unknown image is an image of an apple or an orange. However, if the picture is of a cat, the multi-class classifier can not tell us this. One-class classification on the other hand (see, e.g.,~\cite{10.1007/978-3-540-89378-3_32}), trains only seeing positive targets, not exposed to any negative samples. An example for such classification is to train the one-class on images of apples. Then when it sees a tomato or an orange, it classifies it as ''not an apple'', as long as the tomato or orange are significantly distinct from than the apple. Such sn approach is helpful when we do not have access to a large collection of representative negative data. This is the case with mmWave images, where negative captures are harder to get. Additional advantage of one-class classifier is the ability to enroll into the system without usage of remote dataset which contains the negative data. 

\begin{figure}[!t]
\centering
\includegraphics[width=6.0cm]{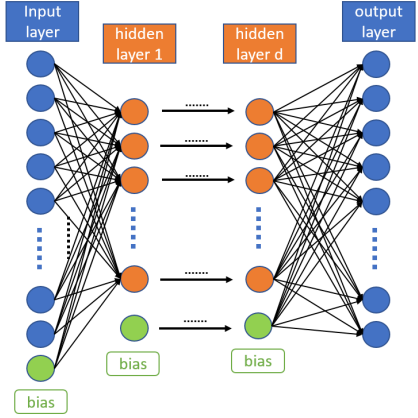}
\caption{Deep autoencoder} \label{fig:deepAutoencoderScheme}
\end{figure}

%\subsection{Autoencoder Using Neural Network} \label{sec:autoencoder}
An autoencoder is a technique to build a one-class classifier in which the input is encoded into a compressed representation through an encoder. The compressed representation is then decoded back into an output. A good autoencoder is one in which the input and the output are similar to each other in a minimum mean square error (MSE) sense. A common implementation of the encoder and the decoder in the autoencoder is composed of a feed-forward artificial neural network with the same input and target output as seen in Fig. \ref{fig:deepAutoencoderScheme} \cite{Hinton504},\cite{vincent10a}. A small hidden layer in an autencoder network creates an information bottleneck, forcing the network to compress the data into a low-dimensional representation. For a simple autoencoder with a single hidden layer, the vector of the hidden unit activities, $\it {h}$, is given by
\begin{equation}
\label{eq:hiddenLayer1}
\mathit{h} = \textit{f}(\mathit{W_{e}\cdot a + bias_{e}})
\end{equation}
where $\it {f}$ is the activation function (we use the logistic sigmoid function in this work), $\it {W_{e}}$ is a parameter matrix, and $\it {bias_{e}}$ is a vector of bias parameters. The hidden representation of the data is then mapped back into the space of $\it {a}$ using the decoding function:
\begin{equation}
\label{eq:outputLayer1}
\mathit{\hat{a}} = \textit{f}(\mathit{W_{d}\cdot h + bias_{d}})
\end{equation}
where $\it {W_{d}}$ is the decoding matrix and $\it {bias_{d}}$ a vector of bias parameters. We learn the parameters of the autoencoder by performing stochastic gradient descent to minimize the reconstruction error which is the MSE between $\it {a}$ and $\it {\hat{a}}$ (mean over the dimension of $\it {a}$):
\begin{equation}
\label{eq:errorFunc1}
\mathit{MSE(a, \hat{a}) = \left \| a -   \hat{a}\right\|_{2}^{2} = \left \| a -   f(\mathit{W_{d}\cdot h + bias_{d}})\right\|_{2}^{2}}.
\end{equation}
When the hidden layer has fewer dimensions than $\it {a}$, the autoencoder learns a compressed representation of the training data. Non-linear hidden units allow the autoencoder to learn more complex encoding functions, as do additional hidden layers.

\subsection{Results}
\begin{figure}[!t]
\centering
\includegraphics[width=6.0cm]{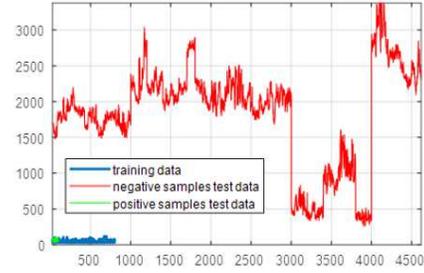}
\caption{MSE results for an autoencoder trained on a specific person, as the ordinate. The abscissa stands for the captured frame index.} \label{fig:mse_res}
\end{figure}

\begin{figure}[!t]
\centering
\includegraphics[width=6.0cm]{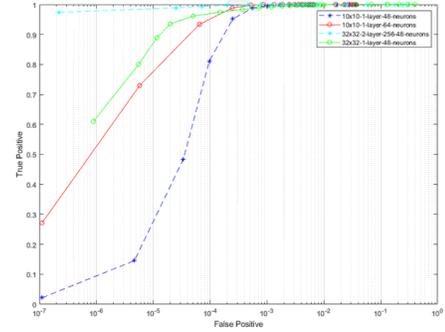}
\caption{ROC for several tested networks and inputs. The 32 transmit antennas with 32 receive antennas with one (solid circles) or two (dashed stars) hidden layers are the two upper curves. 10 transmit and 10 receive antennas are the two lower curves, where 48 wide layer is clearly insufficient when compared to 64 neurons.} \label{fig:roc}
\end{figure}

The MSE as a result of one of the faces is shown in Fig. \ref{fig:mse_res} (MSE is over the 6K dimensions of the captured frame). The MSE quantifies how well the autoencoder encoded the trained data. We see in this figure that the training data from a specific person resulted with MSE of 60. When running the test data (due to the low number of data samples we used only $10\%$ as test data) we get the same MSE of around 60. When we run the trained autoencoder over the negative data (other faces), we got MSE of no less than 250, where 78\% of the captured frames resulted with an MSE of over 1500. This is enough separation to distinguish between the different faces in the captured dataset. 
We next tuned a threshold value on the MSE to determine the region of convergence (ROC) by running the trained autoencoders over the entire ensemble of captured faces. This is shown in Fig. \ref{fig:roc}. For the two hidden layers scheme, we observe excellent results, with false negative of less than 2\%, we get false positive $<10^{-6}$. This is indicative of a strong distinctiveness between the radar signatures of different faces and the high correlation between samples of the same person. In Fig. \ref{fig:roc} we checked several configurations. We tested what happens when we reduce the dimensionality of the radar signature 10 times by taking only 10 antennas out of the 32. It is seen that such reduction significantly reduces the distinctiveness between the faces. We also checked what is the right configuration for the autoencoder network, and noticed an improvement once we increase the number of neurons in the hidden layer and when an additional hidden layer was added.  

\section{Conclusions} \label{sec:conclus} 
A dataset of face signatures as captured with a mmWave radar is introduced. This dataset contains the capturing of about 200 faces of different people. The faces captured include two genders and multiple ages. In addition, the set includes people with and without eyeglasses and beards.
One of the key findings of our research is that the dataset shows distinctiveness between the faces of different persons. Moreover, is our study shows that there is a correlation between different captures of the same face. In our study we trained a deep autoencoders based on neural networks. With those autoencoders, we demonstrated promising results indicating the potential of using mmWave signature as an additional modality for facial verification/recognition. mmWave can penetrate through fabrics and hair and thus can provide a more robust, reliable and secure verification. By re-using the commercially available communication chipset, the solution can be low cost as well. 

\bibliographystyle{ieeetr}
\bibliography{mybibliography}

\begin{thebibliography}{10}

\bibitem{wayman2005introduction}
J.~Wayman, A.~Jain, D.~Maltoni, and D.~Maio, ``An introduction to biometric
  authentication systems,'' in {\em Biometric Systems}, pp.~1--20, Springer,
  2005.

\bibitem{1467368}
P.~J. Phillips, P.~J. Flynn, T.~Scruggs, K.~W. Bowyer, J.~Chang, K.~Hoffman,
  J.~Marques, J.~Min, and W.~Worek, ``Overview of the face recognition grand
  challenge,'' in {\em 2005 IEEE Computer Society Conference on Computer Vision
  and Pattern Recognition (CVPR'05)}, vol.~1, pp.~947--954 vol. 1, June 2005.

\bibitem{abate20072d}
A.~F. Abate, M.~Nappi, D.~Riccio, and G.~Sabatino, ``{2D and 3D} face
  recognition: A survey,'' {\em Pattern recognition letters}, vol.~28, no.~14,
  pp.~1885--1906, 2007.

\bibitem{tran2017regressing}
A.~T. Tran, T.~Hassner, I.~Masi, and G.~Medioni, ``Regressing robust and
  discriminative {3D} morphable models with a very deep neural network,'' in
  {\em Computer Vision and Pattern Recognition (CVPR), 2017 IEEE Conference
  on}, pp.~1493--1502, IEEE, 2017.

\bibitem{3DAssistedFaceRecognition2005}
J.~Kittler, A.~Hilton, M.~Hamouz, and J.~Illingworth, ``3d assisted face
  recognition: A survey of 3d imaging, modelling and recognition approachest,''
  in {\em 2005 IEEE Computer Society Conference on Computer Vision and Pattern
  Recognition (CVPR'05) - Workshops(CVPRW)}, vol.~00, p.~114, 06 2005.

\bibitem{Bakirman17}
T.~Bakirman, M.~U. Gumusay, H.~C. Reis, M.~O. Selbesoglu, S.~Yosmaoglu, M.~C.
  Yaras, D.~Z. Seker, and B.~Bayram, ``Comparison of low cost 3d structured
  light scanners for face modeling,'' {\em Appl. Opt.}, vol.~56, pp.~985--992,
  Feb 2017.

\bibitem{luna2017robust}
C.~A. Luna, C.~Losada-Gutierrez, D.~Fuentes-Jimenez, A.~Fernandez-Rincon,
  M.~Mazo, and J.~Macias-Guarasa, ``Robust people detection using depth
  information from an overhead time-of-flight camera,'' {\em Expert Systems
  with Applications}, vol.~71, pp.~240--256, 2017.

\bibitem{ABIResearch60Ghz}
{ABI Research}, ``802.11ad will vastly enhance {Wi-Fi} the importance of the 60
  {GHz} band to {Wi-Fi}'s continued evolution,'' tech. rep., April 2016.

\bibitem{smulders2002exploiting}
P.~Smulders, ``Exploiting the 60 {GHz} band for local wireless multimedia
  access: prospects and future directions,'' {\em IEEE communications
  magazine}, vol.~40, no.~1, pp.~140--147, 2002.

\bibitem{molchanov2015short}
P.~Molchanov, S.~Gupta, K.~Kim, and K.~Pulli, ``Short-range {FMCW} monopulse
  radar for hand-gesture sensing,'' in {\em Radar Conference (RadarCon), 2015
  IEEE}, pp.~1491--1496, IEEE, 2015.

\bibitem{lien2016soli}
J.~Lien, N.~Gillian, M.~E. Karagozler, P.~Amihood, C.~Schwesig, E.~Olson,
  H.~Raja, and I.~Poupyrev, ``Soli: Ubiquitous gesture sensing with millimeter
  wave radar,'' {\em ACM Transactions on Graphics (TOG)}, vol.~35, no.~4,
  p.~142, 2016.

\bibitem{yeo2016radarcat}
H.-S. Yeo, G.~Flamich, P.~Schrempf, D.~Harris-Birtill, and A.~Quigley,
  ``Radarcat: Radar categorization for input \& interaction,'' in {\em
  Proceedings of the 29th Annual Symposium on User Interface Software and
  Technology}, pp.~833--841, ACM, 2016.

\bibitem{harmer2012review}
S.~Harmer, N.~Bowring, D.~Andrews, N.~Rezgui, M.~Southgate, and S.~Smith, ``A
  review of nonimaging stand-off concealed threat detection with
  millimeter-wave radar [application notes],'' {\em IEEE Microwave magazine},
  vol.~13, no.~1, pp.~160--167, 2012.

\bibitem{sakamoto2016fast}
T.~Sakamoto, T.~Sato, P.~Aubry, and A.~Yarovoy, ``Fast imaging method for
  security systems using ultrawideband radar,'' {\em IEEE Transactions on
  Aerospace and Electronic Systems}, vol.~52, no.~2, pp.~658--670, 2016.

\bibitem{gonzalez2016millimeter}
B.~Gonzalez-Valdes, Y.~{\'A}lvarez, Y.~Rodriguez-Vaqueiro,
  A.~Arboleya-Arboleya, A.~Garc{\'\i}a-Pino, C.~M. Rappaport, F.~Las-Heras, and
  J.~A. Martinez-Lorenzo, ``Millimeter wave imaging architecture for
  on-the-move whole body imaging,'' {\em IEEE Transactions on Antennas and
  Propagation}, vol.~64, no.~6, pp.~2328--2338, 2016.

\bibitem{lopez2018using}
S.~L{\'o}pez-Tapia, R.~Molina, and N.~P. de~la Blanca, ``Using machine learning
  to detect and localize concealed objects in passive millimeter-wave images,''
  {\em Engineering Applications of Artificial Intelligence}, vol.~67,
  pp.~81--90, 2018.

\bibitem{yujiri2003passive}
L.~Yujiri, M.~Shoucri, and P.~Moffa, ``Passive millimeter wave imaging,'' {\em
  IEEE microwave magazine}, vol.~4, no.~3, pp.~39--50, 2003.

\bibitem{ahmed2012advanced}
S.~S. Ahmed, A.~Schiessl, F.~Gumbmann, M.~Tiebout, S.~Methfessel, and
  L.~Schmidt, ``Advanced microwave imaging,'' {\em IEEE microwave magazine},
  vol.~13, no.~6, pp.~26--43, 2012.

\bibitem{noujeim2014compact}
K.~Noujeim, G.~Malysa, A.~Babveyh, and A.~Arbabian, ``A compact
  nonlinear-transmission-line-based mm-wave {SFCW} imaging radar,'' in {\em
  Microwave Conference (EuMC), 2014 44th European}, pp.~1766--1769, IEEE, 2014.

\bibitem{feger2013low}
R.~Feger, A.~Fischer, and A.~Stelzer, ``Low-cost implementation of a millimeter
  wave imaging system operating in w-band,'' in {\em Microwave Symposium Digest
  (IMS), 2013 IEEE MTT-S International}, pp.~1--4, IEEE, 2013.

\bibitem{oppelt2017mimo}
D.~Oppelt, J.~Adametz, J.~Groh, O.~Goertz, and M.~Vossiek, ``{MIMO-SAR} based
  millimeter-wave imaging for contactless assessment of burned skin,'' in {\em
  Microwave Symposium (IMS), 2017 IEEE MTT-S International}, pp.~1383--1386,
  IEEE, 2017.

\bibitem{haderer2014comparison}
H.~Haderer, R.~Feger, and A.~Stelzer, ``A comparison of phase-coded {CW} radar
  modulation schemes for integrated radar sensors,'' in {\em Microwave
  Conference (EuMC), 2014 44th European}, pp.~1896--1899, IEEE, 2014.

\bibitem{levanon2017complementary}
N.~Levanon, I.~Cohen, and P.~Itkin, ``Complementary pair radar
  waveforms--evaluating and mitigating some drawbacks,'' {\em IEEE Aerospace
  and Electronic Systems Magazine}, vol.~32, no.~3, pp.~40--50, 2017.

\bibitem{pezeshki2008doppler}
A.~Pezeshki, A.~R. Calderbank, W.~Moran, and S.~D. Howard, ``Doppler resilient
  {Golay} complementary waveforms,'' {\em IEEE Transactions on Information
  Theory}, vol.~54, no.~9, pp.~4254--4266, 2008.

\bibitem{kumari2015investigating}
P.~Kumari, N.~Gonzalez-Prelcic, and R.~W. Heath, ``Investigating the ieee
  802.11 ad standard for millimeter wave automotive radar,'' in {\em Vehicular
  Technology Conference (VTC Fall), 2015 IEEE 82nd}, pp.~1--5, IEEE, 2015.

\bibitem{kumari2017ieee}
P.~Kumari, J.~Choi, N.~G. Prelcic, and R.~W. Heath, ``Ieee 802.11 ad-based
  radar: An approach to joint vehicular communication-radar system,'' {\em IEEE
  Transactions on Vehicular Technology}, 2017.

\bibitem{grossi2018opportunistic}
E.~Grossi, M.~Lops, L.~Venturino, and A.~Zappone, ``Opportunistic radar in ieee
  802.11 ad networks,'' {\em IEEE Transactions on Signal Processing}, 2018.

\bibitem{wenger1998automotive}
J.~Wenger, ``Automotive mm-wave radar: Status and trends in system design and
  technology,'' 1998.

\bibitem{hasch2012millimeter}
J.~Hasch, E.~Topak, R.~Schnabel, T.~Zwick, R.~Weigel, and C.~Waldschmidt,
  ``Millimeter-wave technology for automotive radar sensors in the 77 {GHz}
  frequency band,'' {\em IEEE Transactions on Microwave Theory and Techniques},
  vol.~60, no.~3, pp.~845--860, 2012.

\bibitem{nitsche2014ieee}
T.~Nitsche, C.~Cordeiro, A.~B. Flores, E.~W. Knightly, E.~Perahia, and J.~C.
  Widmer, ``Ieee 802.11 ad: directional 60 {GHz} communication for
  multi-gigabit-per-second {Wi-Fi},'' {\em IEEE Communications Magazine},
  vol.~52, no.~12, pp.~132--141, 2014.

\bibitem{10.1007/978-3-540-89378-3_32}
K.~Hempstalk and E.~Frank, ``Discriminating against new classes: One-class
  versus multi-class classification,'' in {\em AI 2008: Advances in Artificial
  Intelligence} (W.~Wobcke and M.~Zhang, eds.), pp.~325--336, 2008.

\bibitem{Hinton504}
G.~E. Hinton and R.~R. Salakhutdinov, ``Reducing the dimensionality of data
  with neural networks,'' {\em Science}, vol.~313, no.~5786, pp.~504--507,
  2006.

\bibitem{vincent10a}
P.~Vincent, H.~Larochelle, I.~Lajoie, Y.~Bengio, and P.~Manzagol, ``Stacked
  denoising autoencoders: Learning useful representations in a deep network
  with a local denoising criterion,'' {\em Journal of Machine Learning
  Research}, vol.~11, pp.~3371−--3408, 2010.

\end{thebibliography}

\end{document}